\documentclass[12pt]{article}
\usepackage{amssymb,amsmath,latexsym}
\usepackage[utf8]{inputenc}
\usepackage[T1]{fontenc}
\usepackage[english]{babel} %francais
\usepackage{wrapfig}
\usepackage{subcaption}
\usepackage{graphicx}
\usepackage{caption}
\usepackage{color}
\usepackage{epstopdf}
\usepackage{hyperref} % pour equations et reference colors
\DeclareMathOperator{\e}{\text{e}}

\DeclareMathOperator{\diag}{\text{diag}}
\newcommand{\ic}{\text{i}}

\begin{document}
\title{Creating mirror-mirror quantum correlations in optomechanics}

\author{
M. Amazioug$^{1,2}${\footnote{%
email: \textsf{amazioug@gmail.com}}}, B. Maroufi$^{2,3}${\footnote{%
email: \textsf{mbouchra@hotmail.com}}}~and~M. Daoud$^{4,5}${\footnote{%
email: \textsf{m\_daoud@hotmail.com}}}\\
{\small $^{1}$ Department of Physics, Ecole Normale Sup\'erieure (ENS), Mohammed V University in Rabat, Morocco } \\
{\small $^{2}$ LPHE-MS, Department of Physics, Faculty of Sciences, Mohammed
V University, Rabat, Morocco } \\
{\small $^{3}$ LPMC, Department of Physics, Faculty of Sciences Ben M'Sik, University Hassan II, Casablanca, Morocco } \\
{\small $^{4}$ Department of Physics, Faculty of Sciences, University Ibn
Tofail, K\'{e}nitra, Morocco } \\
{\small $^{5}$ Abdus Salam International Centre for Theoretical Physics,
Strada Costiera, 11, 34151 Trieste, Italy
}}
\date{}
\maketitle

\textbf{Abstract:} We study the transfer of quantum correlations between two movable mirrors of two Fabry-P\'erot cavities separated via broadband squeezed light and coupled via photon hopping process. We investigate the transfer of quantum correlations from EPR entangled squeezed light to the movable mirrors. We show that Gaussian quantum steering remains lower than entanglement. We employ Gaussian quantum steering to characterize the steerability between the two mechanical modes. The logarithmic negativity is used as the witness of quantum entanglement and Gaussian quantum discord gives the measure of all non classical correlations including entanglement. We conclude that the transfer of quantum correlations is optimal for a strong optomechanical coupling and decreases with the thermal effects. We also conclude that steering, entanglement and discord are directly related to photon hopping coupling and the squeezing parameter.


\section{Introduction}

Cavity optomechanical system constitutes a good platform to study the interaction between light and mechanical oscillators \cite{Aspelmeyer2014,JQLiao2011}. Indeed, this interaction is generated by the radiation pressure between the movable mirror and the cavity field of a cavity of Fabry-P\'erot type \cite{Aspelmeyer2012}. Cavity optomechanics have attracted a special attention during the last two decades, especially
in the area of quantum information processing to implement various interesting quantum tasks such
cooling the mechanical mode to their quantum ground states \cite{JTeufel2011,SMachnes2012,JChan2011,MBhattacharya2007}, generating a mechanical quantum superposition \cite{JQLiao2016}, realizing entanglement between mechanical and optical modes in steady state \cite{SMancini2002,MJHartmann2008,DVitali2007}, quantum measurements precision \cite{JQZhang2012,ZXLiu2017,HXiong2017110,HXiong201742} and gravitational-wave detectors \cite{CMCaves1980,AAbramovici1992,VBraginsky2002}. Several studies have considered the determination  of the quantum correlations (QCs) between
the optical and the mechanical modes in optomechanical systems
\cite{MPinard2005,SHuang2009,EASete2014,SMancini2002,CGenes2008,JLi2017,MAsjad2016,
amaziougEPJD2018,Amazioug1,Amazioug2,SChakrabortyd2018,Amazioug3,SBougouffa2019}. This interest is
mainly motivated by the fact the encoding quantum information is easy to implement in continuous variable (CV) states   in comparison with discrete variable states \cite{AMann1995}. \\

In
the context of optomechanics, the information is encoded in Gaussian states and many authors considered the characterization of the nature of the existing
correlations shared by the optical and mechanical modes \cite{IKogias2015,SMancini2002,RSimon2000,LMDuan2000,GVidal2002,GAdesso2004,PMarian2008,GAdesso2010,PGiorda2010}. Different
 quantifiers were employed such as quantum entanglement (QE),  quantum steering (QS) and quantum discord (QD) to show the role of the quantum correlations as a fundamental resource
for several protocols in quantum information \cite{PSkrzypczyk2014,YChen2014,IKogias2015,MPiani2015,MCLi2015,CHBennett1996,GVidal2002,RHorodecki2009,HOllivier2001,GAdesso2010,PGiorda2010,KModi2012,
SLuo2008,BLi2011,MAli2010}. Quantum entanglement
 were proposed by Schr\"odinger in framework of the EPR paradox \cite{EPR1935,schrodinger1935}. Quantum steering
 is used to quantify how much the two entangled bipartite states are steerable. This quantifier exhibits a the
 asymmetric property between two entangled observers (Alice and Bob). In this sense, Alice can change (i.e. 'steer') the Bob states by exploiting their shared entanglement \cite{IKogias2015}.
 Entanglement  emerges as an important ingredient at the heart of quantum information and processing, such as quantum teleportation \cite{Bennett1993}, superdense coding \cite{Bennett1992}, telecloning \cite{Scarani2005} and quantum cryptography \cite{Ekert1991}.  Clearly, to exploit the quantum features of non classical correlations in quantum information processing, it is necessary
 to tackle the decoherence effects \cite{Zurek2003}. The coupling between the  system and its own environment
 induces the loss of quantum correlations and in some special this can occur suddenly.
This is called the entanglement sudden death (ESD) \cite{AAlQasimi2008}. This phenomenon emerges when the entangled multipartite quantum system is placed in Markovian environments \cite{Yu2004,Yu2006Opt,Yu2006PRL,Yu2009,Almeida2007}. It must be noticed that the entanglement sudden death can be exhibited in multi-qubit systems evolving in
non Makovian environment (see for instance \cite{Park}). Furthermore, it has been shown that under some special circumstances, the entanglement can
be generated. This generation is termed in the literature entanglement sudden birth (ESB) \cite{ZFicek2006}.  The quantum discord is another quantum correlation
quantifier introduced to go beyond entanglement. It is essentially used in quantifying the quantumness in multi-partite systems in view of its robustness against decoherence in comparison with entanglement(see for instance  \cite{GFernando2011} and references therein).\\

Besides the characterization of the nature and the evaluation of the amount of quantum correlations between
the optical and the mechanical degrees of freedom in optomechanical systems, some recent studies were dedicated to
the transfer of the non classical correlations from optical to mechanical modes \cite{SHuang2009,EASete2014,amaziougEPJD2018,Jamal2015,JamalPRA,SBougouffa2017}.  It
is commonly accepted that this quantum correlations transfer provides a potential tool to exploit the quantum information encoded in mechanical modes which can be more resilient against
decoherence effects. The enhancement of this transfer is crucial. In this sense, the photon hopping process, through which two optomechanical cavities can be
coupled, was considered in  \cite{SBougouffa2017} to show its role in generating
 entanglement between mechanical oscillators and in enhancing the amount of the entanglement transfer from optical to mechanical modes. In connection
 with this transfer process, it has been recently shown that the production of entanglement and its transfer depends strongly on photon hopping strength \cite{SBougouffa2019}.\\

This work aims to contribute to the investigation of the quantum correlation transfer by exploiting the photon hopping coupling in optomechanical systems. Furthermore,
we shall study the amount of quantum correlations distributed between the different components of the optomechanical systems by employing quantum steering and quantum discord. This
will extend the analysis previously initiated in the works \cite{SBougouffa2017,SBougouffa2019} where  logarithmic negativity is used to quantify non classical correlations. We also
consider the thermal effects on converting the quantum correlations from photons (optical modes) to phonons (mechanical modes).  Our analysis are conducted
in the framework of  rotating wave approximation (RWA). The system under consideration consists of two cavities separated in space where each cavity has a movable end-mirror. These two cavities are coupled to a two-mode squeezed light generated from spontaneous parametric down-conversion and coupled by photon hopping process as shown in Figure \ref{schema}. The optical modes are entangled when the squeezed parameter $r$ is no zero \cite{Jamal2015,amaziougEPJD2018}. In this picture we shall analyse the quantum correlations transfer fom entangled optical modes to mechanical modes associat to the movable mirrors. We use Gaussian quantum steering \cite{IKogias2015} to characterize steerability of the two mechanical modes. We use the logarithmic negativity \cite{GVidal2002,GAdesso2004} to quantify the amount of quantum entanglement between the two mechanical modes for mixed Gaussian states. Gaussian quantum discord is used to quantify all quantum correlations existing between the two movable mirrors, i.e. even if they are not entangled. We show that the three quantum correlations indicators, namely QS, QE and QD, are affected by the thermal effects of baths of two movable mirrors and depend of the optomechanical cooperativity.\\

The article is organized as follows. In Section 2, we give the model of the system under study, the Hamiltonian governing the optomechanical system under consideration and quantum nonlinear Langevin equations (QLEs) for mechanical and optical modes. In section 3, we linearize the QLEs and derive the quantum equations that describe the steady state of the system. In Section 4, we evaluate the covariance matrix of steady state for CV states in framework of RWA. Section 5 is devoted to the study of quantum correlations between the two mechanical modes using Gaussian quantum steering, logarithmic negativity, Gaussian quantum discord. The obtained results are discussed.
Concluding remarks close this paper.

\section{Model}

We first describe the optomechanical system under consideration and give the Hamiltonian governing the evolution of the
corresponding subsystems. The global system is composed of two Fabry-P\'erot cavities each cavity has a movable end-mirror.

\begin{figure}[!htb]
\centering
\includegraphics[width=16cm,height=5cm]{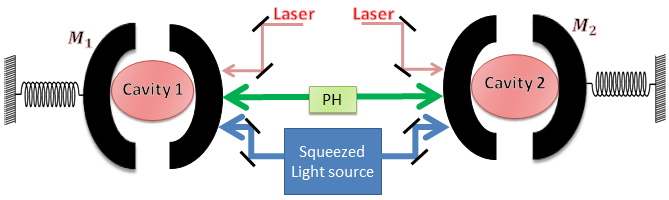}
\caption{Schematics of two Fabry-P\'erot cavities 1(2) separated in space and coupled to a two-mode squeezed light from spontaneous parametric down-conversion (SPDC) and coupled by photon hopping process (PH) with coupling strength $\lambda$. Each cavity is pumped by coherent laser source with amplitude $E_{j}$.}
\label{schema}
\end{figure}

The two cavities are coupled by photon hopping process and the output of squeezed light source. Each cavity is pumped by coherent laser source. Moreover, the movable mirror (the mass and the frequency of the $j^{th}$ movable mirror $M_j(j=1,2)$ are respectively $m_j$ and $\omega_{M_{j}}$) is coupled to the photons inside the cavity via radiation pressure (Fig. \ref{schema}). The coupling rate $g_j=\frac{\omega_{c_{j}}}{L_j}\sqrt{\frac{\hbar}{m_j\omega{_M}_j}}$, with $j^{th}$ cavity length is $L_j$ \cite{Aspelmeyer2014}.\\
The Hamiltonian of the system is in a rotating frame rotating with $\omega_{L_j}$ is given by writes as :
\begin{equation}\label{eq:1}
\mathcal{H}=\sum_{j=1}^{2}\left[\hbar\omega_{M_{j}}\left(b^{+}_{j}b_{j}+\frac{1}{2}\right)-\hbar\Delta_{j}c^{+}_{j}c_{j}-\hbar g_{j}c^{+}_{j}c_{j}(b^{+}_{j}+b_{j}) +\hbar(c^{+}_{j}E_{j}\e^{\ic\varphi_j}+c_{j}E_{j}\e^{-\ic\varphi_{j}})\right]-\hbar\lambda (c_1^+ c_2+c_2^+ c_1)
\end{equation}
where $\kappa_j$ and $\omega_{c_{j}}$ are respectively the damping rate and the frequency of $j^{th}$ cavity, $E_j=\sqrt{\frac{2\kappa_j \wp_j}{\hbar\omega{_L}_j}}$ and $\varphi_j$ ($j=1,2$) are respectively the amplitude and the phase of the input coherent laser field. $\wp_j$ is the drive pump power and $\omega{_L}_j$ is the frequency of $j^{th}$ input field. The operators of annihilation and creation of $j^{th}$ movable mirrors are respectively $b_j$ and $b^+_j$. They satisfy the commutation rules $[b_j, b_j^+]=1$ ($j=1,2$) and the annihilation and creation operators of the $j^{th}$ cavity mode are respectively $c_j$ and $c_j^+$ with $[c_j,c_j^+]=1$ ($j=1,2$). The non linear quantum Langevin equations of the mechanical and optical modes are
\begin{equation} \label{eq:2}
\frac{\partial}{\partial t} b_{j}=-\left(\ic\omega_{M_{j}}+\frac{\gamma_j}{2}\right)b_{j}+\ic g_{j}c^{+}_{j}c_{j} + \sqrt{\gamma_{j}}b^{in}_{j}
\end{equation}
\begin{equation} \label{eq:3}
\frac{\partial}{\partial t} c_{j}=-\left(\frac{\kappa_{j}}{2}-\ic\Delta_{j}\right)c_{j}+\ic g_{j}c_{j}(b^{+}_{j} + b_{j})-\ic E_{j}\e^{\ic\varphi_{j}}+\ic\lambda c_s+\sqrt{\kappa_{j}}c^{in}_{j} ; j\neq s
\end{equation}
with the $j^{th}$ mechanical damping rate is $\gamma_j$ and the $j^{th}$ laser detuning is $\Delta_{j}=\omega_{L_{j}} - \omega_{c_{j}}$. We assume that $ \omega_{M_j} \gg \gamma_j $ and $\Delta_j \gg \kappa_j $ so that we can apply the rotating wave approximation. The $j^{th}$ noise operator describing the coupling between the mechanical mode and its own environment is $b^{in}_{j}$  and the $j^{th}$ squeezed vacuum operator is denoted by $c^{in}_{j}$. One can assume that the mechanical baths are Markovian when the mechanical quality factor is very large ($Q\gg 1$). The non-zero correlation factors \cite{VGiovannetti2001,CWGardiner2000} are :
\begin{equation} \label{eq:4}
\langle b^{in}_{j}(t)b^{in+}_{j}(t')\rangle = (n_{th_j}+1)\delta(t-t')
\end{equation}
\begin{equation} \label{eq:5}
\langle b^{in+}_{j}(t)b^{in}_{j}(t')\rangle = n_{th_j}\delta(t-t')
\end{equation}
$n_{th_j}=\left[\exp\left(\frac{\hbar\omega_{M_{j}}}{k_B T_j}\right)-1\right]^{-1}$ is the thermal baths phonons numbers in the $j^{th}$ cavity, with the Boltzmann constant $k_B$. The non-zero correlations properties of the squeezed vacuum operators $c^{in}_{j}$ and $c^{in+}_{j}$ are given by \cite{CWGardiner1986}
\begin{equation} \label{eq:6}
\langle c^{in}_{j}(t)c^{in+}_{j}(t')\rangle = (\mathcal{N}+1)\delta(t-t')
\end{equation}
\begin{equation} \label{eq:7}
\langle c^{in+}_{j}(t)c^{in}_{j}(t')\rangle = \mathcal{N}\delta(t-t')
\end{equation}
\begin{equation} \label{eq:8}
\langle c^{in}_{j}(t)c^{in}_{j'}(t')\rangle =  \mathcal{M}\e^{-\ic\omega_{M}( t+t')}\delta(t-t')\quad;\quad j\neq j'
\end{equation}
\begin{equation} \label{eq:9}
\langle c^{in+}_{j}(t)c^{in+}_{j'}(t')\rangle =  \mathcal{M}\e^{\ic\omega_{M}( t+t')}\delta(t-t')\quad;\quad j\neq j'
\end{equation}
where $\mathcal{N}=\sinh^2 r$, $\mathcal{M}=\sinh r\cosh r$, $r$ is the squeezing parameter characterizing the squeezed light (we have considered $\omega_{M_1}=\omega_{M_2}=\omega_M$).

\section{Linearization of quantum Langevin equations}

The equations (\ref{eq:2}) and (\ref{eq:3}) in general are difficult to solve solvable analytically \cite{EASete2014}. To simplify this, we adopt the linearization scheme \cite{EASete2011}
\begin{equation} \label{eq:10}
b_j=\delta b_j+\bar{b}_j\quad;\quad c_j=\delta c_j+\bar{c}_j
\end{equation}
with $\delta b_j$ and $\delta c_j$ being the fluctuations operators. Using Eqs. (\ref{eq:2}) and (\ref{eq:3}) the steady-state average $\bar{b}_j$ and $\bar{c}_j$ associated with the operators $b_j$ and $c_j$ are
\begin{equation} \label{eq:11}
\bar{c}_1=\frac{\ic\alpha_2 E_1\e^{\ic\varphi_1}+\lambda E_2\e^{\ic\varphi_2}}{\alpha_2\alpha_1+\lambda^2}\quad;\quad \bar{c}_2=\frac{\ic\alpha_1 E_2\e^{\ic\varphi_2}+\lambda E_1\e^{\ic\varphi_1}}{\alpha_2\alpha_1+\lambda^2} \quad;\quad \bar{b}_j=\frac{\ic g_j \mid \bar{c}_j \mid ^2}{\frac{\gamma_j}{2}+\ic\omega_{M_{j}}}; j=1,2
\end{equation}
with $\alpha_j=-\frac{\kappa_j}{2}+\ic\Delta'_j$ and $\Delta'_j=\Delta_j+g_j (\bar{b}_j^* + \bar{b}_j)$. For $\alpha_1=\alpha_2=\alpha$, $E_1=E_2=E$, $\varphi_{1}=\varphi_{2}=\varphi$, one has $\bar{c}=\ic\frac{E\e^{\ic\varphi}}{\frac{\kappa}{2}-\ic(\Delta'+\lambda)}$. For an arbitrary phase of the coherent drives, we use the phase of the input laser to be $\varphi=-\arctan[2(\Delta'+\lambda)/\kappa]$ so that $\bar{c}=\ic\mid \bar{c} \mid$. In this way, the many-photon optomechanical coupling inside the $j^{th}$ cavity is \cite{Aspelmeyer2014} :
\begin{equation}
\mathcal{G}_j=g\mid \bar{c} \mid=\frac{\omega_{c}}{L}\sqrt{\frac{2\kappa\wp}{m\omega_{M}\omega_{L}[(\Delta'+\lambda)^2 + \kappa^2/4]}}
\end{equation}

The equations (\ref{eq:2}) and (\ref{eq:3}) give
\begin{equation} \label{eq:12}
\frac{\partial}{\partial t}\delta b_{j}=-\left(\ic\omega_{M_{j}}+\frac{\gamma}{2}\right)\delta b_{j}+\mathcal{G}_{j}(\delta c_{j}-\delta c^+_{j}) + \sqrt{\gamma}b^{in}_{j}
\end{equation}

\begin{equation} \label{eq:13}
\frac{\partial}{\partial t}\delta c_{j}=-\left(\frac{\kappa}{2}-\ic\Delta'_{j}\right)\delta c_{j}-\mathcal{G}_{j}(\delta b^{+}_{j} + \delta b_{j})+\ic\lambda \delta c_s+\sqrt{\kappa}c^{in}_{j} ; j\neq s
\end{equation}
with the effective cavity detuning $\Delta'_j=\Delta_j+g_j(\bar{b}_j+\bar{b}_j^*)$ is related to the displacement of the mirrors generated by the radiation pressure force.

We use the notations $\delta c_j(t)=\delta \widetilde{c}_j (t)\e^{\ic\Delta'_j t}$, $\delta b_j (t)=\delta \widetilde{b}_j(t)\e^{-\ic\omega_{M_j} t}$, $\widetilde {c}^{in}_{j}=\e^{-\ic\Delta'_j t}c^{in}_{j}$ and $\widetilde {b}^{in}_{j}=\e^{\ic\omega_{M_j} t}b^{in}_{j}$.

Using the rotating wave approximation (RWA) \cite{Aspelmeyer2014,YDWang2015} (i.e. $\omega_{M_j}\gg\kappa$ ; $j=1,2$), the effective cavity detuning is reduced to  $\Delta'_j\approx\Delta_j$, and we can neglect the rotating terms at $\pm2\omega_{M_j}$. If the cavity is driven at the red sideband ($\Delta'_j=-\omega_{M_j}$), the equations $(\ref{eq:12})$ and $(\ref{eq:13})$ become
\begin{equation} \label{eq:14}
   \delta \dot{\widetilde b}_j=-\frac{\gamma}{2} \delta \widetilde{b}_j + \mathcal{G}_j\delta \widetilde{c}_j +\sqrt{\gamma}\widetilde {b}^{in}_{j} ,
\end{equation}
\begin{equation} \label{eq:15}
   \delta \dot{\widetilde c}_j=-\frac{\kappa}{2}\delta \widetilde{c}_j-\mathcal{G}_j\delta \widetilde{b}_j+\ic\lambda \delta\widetilde{c}_s +\sqrt{\kappa}\widetilde {c}^{in}_{j}, j\neq s
\end{equation}

\section{Covariance matrix of the steady state}

We assume that  the two coherent sources have the same strength so that the state of two cavities are symmetric. Using this fact and
the set of linear quantum Langevin's equations (\ref{eq:14}) and (\ref{eq:15}) of mechanical and optical modes, we shall determine the covariance matrix
of the steady state needed to analyse the pairwise quantum correlations between the different modes of the system.
The temperature of the thermal baths of the two movable mirrors are assumed as before the same $T_1=T_2=T$ ($n_{th_1}=n_{th_2}=n_{th}$) and $m_1=m_2=m$, $\omega_{c_1}=\omega_{c_2}=\omega_c$, $\omega_{M_1}=\omega_{M_2}=\omega_M$, $\kappa_1=\kappa_2=\kappa$ and $\gamma_1=\gamma_2=\gamma$. The EPR-type quadrature operators for the two mechanical and optical modes are
\begin{equation} \label{eq:16}
 \delta \widetilde{q}_{b_j}= \frac{\delta \widetilde{b}_j^++\delta \widetilde{b}_j}{\sqrt{2}}\quad,\quad \delta \widetilde{Y}_{b_j}= \frac{\delta \widetilde{b}_j-\delta \widetilde{b}_j^+}{\ic\sqrt{2}} \quad;\quad j=1 , 2  ,
\end{equation}
\begin{equation} \label{eq:17}
 \delta \widetilde{q}_{c_j}= \frac{\delta \widetilde{c}_j^++\delta \widetilde{c}_j}{\sqrt{2}}\quad,\quad\delta \widetilde{Y}_{c_j}= \frac{\delta \widetilde{c}_j-\delta \widetilde{c}_j^+}{\ic\sqrt{2}} \quad;\quad j=1 , 2  .
\end{equation}
The equations (\ref{eq:14}) and (\ref{eq:15}) rewrite as
\begin{equation} \label{eq:18}
 \delta \dot{\widetilde{q}}_{b_j}=-\frac{\gamma}{2}\delta \widetilde{q}_{b_j}+\mathcal{G}\delta \widetilde{q}_{c_j}+\sqrt{\gamma}\widetilde{q}_{b_j}^{in}\quad;\quad j=1 , 2
\end{equation}
\begin{equation} \label{eq:19}
 \delta \dot{\widetilde{Y}}_{b_j}=-\frac{\gamma}{2}\delta \widetilde{Y}_{b_j}+\mathcal{G}\delta \widetilde{Y}_{c_j}+\sqrt{\gamma}\widetilde{Y}_{b_j}^{in}\quad;\quad j=1 , 2
\end{equation}
\begin{equation} \label{eq:20}
 \delta \dot{\widetilde{q}}_{c_j}=-\frac{\kappa}{2}\delta \widetilde{q}_{c_j}-\mathcal{G}\delta \widetilde{q}_{b_j}-\lambda\delta \widetilde{Y}_{c_s}+\sqrt{\kappa}\widetilde{q}_{c_j}^{in}\quad;\quad j\neq s=1 , 2
\end{equation}
\begin{equation} \label{eq:21}
 \delta \dot{\widetilde{Y}}_{c_j}=-\frac{\kappa}{2}\delta \widetilde{Y}_{c_j}-\mathcal{G}\delta \widetilde{Y}_{b_j}+\lambda\delta \widetilde{q}_{c_s}+\sqrt{\kappa}\widetilde{Y}_{c_j}^{in}\quad;\quad j\neq s=1 , 2
\end{equation}
where
\begin{equation} \label{eq:22}
  \widetilde{q}_{b_j}^{in}= \frac{\widetilde{b}_j^{in+}+ \widetilde{b}^{in}_j}{\sqrt{2}}\quad,\quad\widetilde{Y}_{b_j}^{in}= \frac{ \widetilde{b}^{in}_j- \widetilde{b}^{in+}_j}{\ic\sqrt{2}} \quad;\quad j=1 , 2
\end{equation}
\begin{equation} \label{eq:23}
  \widetilde{q}_{c_j}^{in}= \frac{\widetilde{c}_j^{in+}+ \widetilde{c}^{in}_j}{\sqrt{2}}\quad,\quad\widetilde{Y}_{c_j}^{in}= \frac{ \widetilde{c}^{in}_j- \widetilde{c}^{in+}_j}{\ic\sqrt{2}}\quad;\quad j=1 , 2
\end{equation}
Equations (\ref{eq:18})-(\ref{eq:21}), can be cast in the compact matrix form \cite{AMari2009}
\begin{equation} \label{eq:24}
\dot{\mathcal{u}}(t)=\mathcal{W}\mathcal{u}(t)+\eta(t)
\end{equation}
where $\mathcal{u}^T(t)=(\delta \widetilde{q}_{b_1},\delta \widetilde{Y}_{b_1},\delta \widetilde{q}_{b_2},\delta \widetilde{Y}_{b_2},\delta \widetilde{q}_{c_1},\delta \widetilde{Y}_{c_1},\delta \widetilde{q}_{c_2},\delta \widetilde{Y}_{c_2})$ is the quadrature vector,

$\eta^T(t)=(\sqrt{\gamma}\widetilde{q}_{b_1}^{in},\sqrt{\gamma}\widetilde{Y}_{b_1}^{in},\sqrt{\gamma}\widetilde{q}_{b_2}^{in},\sqrt{\gamma}\widetilde{Y}_{b_2}^{in},\sqrt{\kappa}\widetilde{q}_{c_1}^{in},\sqrt{\kappa}\widetilde{Y}_{c_1}^{in},\sqrt{\kappa}\widetilde{q}_{c_2}^{in},\sqrt{\kappa}\widetilde{Y}_{c_2}^{in})$ is the noise vector and the matrix $\mathcal{W}$ is
\begin{equation} \label{eq:25}
\mathcal{W}=
\begin{pmatrix}
	    -\frac{\gamma}{2} & 0 & 0 & 0 & \mathcal{G} & 0 & 0 & 0 \\
    0 & -\frac{\gamma}{2} & 0 & 0 & 0 & \mathcal{G} & 0 & 0   \\
    0 & 0 & -\frac{\gamma}{2} & 0 & 0 & 0 & \mathcal{G} & 0  \\
    0 & 0 & 0 & -\frac{\gamma}{2} & 0 & 0 & 0 & \mathcal{G} \\
		 -\mathcal{G} & 0 & 0 & 0 & -\frac{\kappa}{2} & 0 & 0 & -\lambda \\
    0 & -\mathcal{G} & 0 & 0 & 0 & -\frac{\kappa}{2} & \lambda & 0   \\
    0 & 0 & -\mathcal{G} & 0 & 0 & -\lambda & -\frac{\kappa}{2} & 0  \\
    0 & 0 & 0 & -\mathcal{G} & \lambda & 0 & 0 & -\frac{\kappa}{2}
\end{pmatrix}
.
\end{equation}

The system is stable when eigenvalues of the drift matrix $\mathcal{W}$ \ref{eq:25} has negative real parts. This corresponds to the so-called Routh-Hurwitz criterion \cite{EXDeJesus1987}. The steady state of the system can be described by Lyapunov equation \cite{VitaliPRL2007,Parks1993} as follows
\begin{equation} \label{eq:26}
 \mathcal{W}\sigma + \sigma \mathcal{W}^T+\mathcal{R}=0   .
 \end{equation}
The explicit expression of the matrix of stationary noise correlations $\mathcal{R}$ are given by $\mathcal{R}_{fh}\delta(t-t') =\frac{1}{2}(\langle\eta_f(t)\eta_h(t')+\eta_h(t')\eta_f(t)\rangle) $ and
\begin{equation} \label{eq:27}
\mathcal{R}=
\begin{pmatrix}
	    \gamma' & 0 & 0 & 0 & 0 & 0 & 0 & 0 \\
    0 & \gamma' & 0 & 0 & 0 & 0 & 0 & 0   \\
    0 & 0 & \gamma' & 0 & 0 & 0 & 0 & 0  \\
    0 & 0 & 0 & \gamma' & 0 & 0 & 0 & 0 \\
		 0 & 0 & 0 & 0 & \kappa' & 0 & \mathcal{M}\kappa & 0 \\
    0 & 0 & 0 & 0 & 0 & \kappa' & 0 & -\mathcal{M}\kappa   \\
    0 & 0 & 0 & 0 & \mathcal{M}\kappa & 0 & \kappa' & 0  \\
    0 & 0 & 0 & 0 & 0 & -\mathcal{M}\kappa & 0 & \kappa'
\end{pmatrix}
\end{equation}
where $\gamma'=\gamma(n_{th}+\frac{1}{2})$ and $\kappa'=\kappa(\mathcal{N}+\frac{1}{2})$. The covariance matrix associated with the two movable mirrors in $(\delta \widetilde{q}_{b_1},\delta \widetilde{Y}_{b_1},\delta \widetilde{q}_{b_2},\delta \widetilde{Y}_{b_2})$ basis writes as
\begin{equation} \label{eq:28}
\sigma_{(m_1m_2)}=
\begin{pmatrix}
	    \sigma_1 & \sigma_{12} & \sigma_{13} & 0 \\
    \sigma_{12} & \sigma_1 & 0 & -\sigma_{13}   \\
    \sigma_{13} & 0 & \sigma_1 & \sigma_{12} \\
    0 & -\sigma_{13} & \sigma_{12} & \sigma_1
\end{pmatrix}
\end{equation}
with
\begin{equation} \label{eq:29}
\sigma_1=\frac{\mathcal{C}(\gamma+\kappa)[\gamma(1+2n_{th})+\kappa^2\cosh(2r)]+(1+2n_{th})(\kappa^2+2\kappa\gamma+\gamma^2+4\kappa^2\xi^2)}{2(\kappa^2+2\kappa\gamma+\gamma^2)(\mathcal{C}+1)+8\kappa^2\xi^2}
\end{equation}
\begin{equation} \label{eq:30}
\sigma_{12}=\frac{\kappa\mathcal{C}\sinh(2r)(\kappa\mathcal{C}+\gamma+2\kappa)\xi}{(\kappa^2+2\kappa\gamma+\gamma^2+4\kappa^2\xi^2)(\mathcal{C}^2+2\mathcal{C}+1+4\xi^2)}
\end{equation}
\begin{equation} \label{eq:31}
\sigma_{13}=\frac{\kappa\mathcal{C}\sinh(2r)[(\gamma+\kappa)(\mathcal{C}+1)-4\kappa\xi^2]}{2(\kappa^2+2\kappa\gamma+\gamma^2+4\kappa^2\xi^2)(\mathcal{C}^2+2\mathcal{C}+1+4\xi^2)}
\end{equation}
where $\mathcal{C}$ is the optomechanical cooperativity \cite{Aspelmeyer2014}
\begin{equation} \label{eq:32}
\mathcal{C}=\frac{4\mathcal{G}^2}{\gamma\kappa}=\frac{8\omega_{c}^2}{m\gamma\omega_{M}\omega_{L} L^2}\frac{\wp}{\kappa^2[1/4+(\xi+\omega_{M}/\kappa)^2]}.
\end{equation}
We notice that in the special case where the photon hopping strength $\xi=\lambda/\kappa=0$ (i.e. $\lambda=0$), the covariance matrix $\sigma_{(m_1m_2)}$ (\ref{eq:28}) coincides with the result reported in \cite{EASete2014,Jamal2015,amaziougEPJD2018,JamalPRA}. The covariance matrix (\ref{eq:28}) can be rewritten as
\begin{equation} \label{eq:33}
\sigma_{(m_1m_2)}=
\begin{pmatrix}
	    \mathcal{X} & \mathcal{Z} \\
    \mathcal{Z}^T & \mathcal{X}
\end{pmatrix}
\end{equation}
where the submatrices $\mathcal{X}=\begin{pmatrix}
	    \sigma_1 & \sigma_{12} \\
    \sigma_{12} & \sigma_1
\end{pmatrix}$ and $\mathcal{Z}=\diag(\sigma_{13},-\sigma_{13})$ represent the covariance matrix $2\times2$ respectively describing the single mode and the quantum correlations between the two movable mirrors.

\section{Quantum correlations}

\subsection{Gaussian quantum steering}

Gaussian quantum steering is the asymmetric property between two entangled observers (the two mechanical modes), Alice (A : mirror $M_1$) and Bob (B : mirror $M_2$). Besides, it can be used to quantify how much the two movable mirrors are steerable. We use the covariance matrix of mechanical mode (see equations Eq. (\ref{eq:28})-(\ref{eq:33})) and Gaussian steering $A\to B$ and $B\to A$ is given by \cite{IKogias2015}
\begin{equation} \label{eq:34}
S^{A \to B}=S^{B \to A}=\max\left[0,\frac{1}{2}\ln{\left(\frac{\det(\mathcal{X})}{4\det\sigma_{(m_1m_2)}}\right)}\right].
\end{equation}
There are two possibilities of steerability between A and B : ($i$) no-way steering if $S^{A \to B}=S^{B \to A}=0$ i.e. Alice can't steer Bob and vice versa even if they are not separable, and ($ii$) two-way steering if $S^{A \to B}=S^{B \to A}>0$, i.e. Alice can steer Bob and vice versa. Indeed, non separable state is not always a steerable state, while a steerable state is always not separable.

\subsection{Quantum entanglement}

The logarithmic negativity $E_N$ is a witness of the entanglement in the bipartite subsystem in CV system. It is defined by \cite{GVidal2002,GAdesso2004}
\begin{equation} \label{eq:35}
	E_{N}=\max[0,-\ln(2\nu^-)]
\end{equation}
with $\nu^-$ being the smallest symplectic eigenvalue of partial transposed covariance matrix $\sigma_{(m_1m_2)}$ (\ref{eq:28}) of two movable mirrors
\begin{equation} \label{eq:36}
\nu^-= \sqrt{\frac{\Delta-\sqrt{\Delta^2-4\det\sigma_{(m_1m_2)}}}{2}}
\end{equation}
where $\Delta$ is given by $\Delta=2(\det \mathcal{X}-\det \mathcal{Z})$.\\
The two movable mirrors are entangled if the condition $\nu^-<1/2$ (i.e. when $E_N>0$) is satisfied.

%%%%%%%%%%%%%%%%%%%%%%%%%%%%%%%%%%%%%%%%%%%%%%%%%%%%%%%%%%%%%%%%%%%%%%%%%%%%%%%%%%%%%%%%%%%%%%%%%%
\subsection{Gaussian quantum discord}
%%%%%%%%%%%%%%%%%%%%%%%%%%%%%%%%%%%%%%%%%%%%%%%%%%%%%%%%%%%%%%%%%%%%%%%%%%%%%%%%%%%%%%%%%%%%%%%%%%

Gaussian quantum discord is used to quantify non classical correlations even if the two movable mirrors are separable. Gaussian quantum discord writes for two movable mirrors as \cite{PGiorda2010}
\begin{equation} \label{eq:37}
\mathcal{D}=f\left(\sqrt{\det \mathcal{X}}\right)-f(\vartheta_+)-f(\vartheta_-)+f(\delta)
\end{equation}
where $f(x)=(x+\frac{1}{2})\ln(x+\frac{1}{2})-(x-\frac{1}{2})\ln(x-\frac{1}{2})$, $\vartheta_+$ and $\vartheta_-$ are the symplectic eigenvalues given by
\begin{equation} \label{eq:38}
\vartheta_{\pm}=\sqrt{\frac{\Delta'\pm\sqrt{\Delta'^2-4\det{\sigma_{(m_1m_2)}}}}{2}}
\end{equation}
with $\Delta'=2(\det \mathcal{X}+\det \mathcal{Z})$ and $\delta$ is
\begin{equation} \label{eq:39}
\delta=\frac{\sqrt{\det \mathcal{X}}+2\det \mathcal{X}+2\det \mathcal{Z}}{1+2\sqrt{\det \mathcal{X}}}.
\end{equation}
The quantum state of two movable mirrors are not separable if $\mathcal{D}>1$. Moreover, if the condition $0\leq\mathcal{D}<1$ is satisfied, the two movable mirrors can be in a separable state or in an entangled state.

\subsection{Resultats and Discusion}

In this section, we will discuss the evolution of quantum correlations under different effects by adopting realistic situation that can be realized experimentally. Parameters we use for our numerical calculations are taken from \cite{SGroblacher2009}: the movable mirrors oscillate with frequency $\omega_M/2\pi =947\times 10^3$ Hz, the mechanical damping rate $\gamma/2\pi= 140$ Hz and having the mass $m$=145 ng. The cavity length and frequency are respectively L=25 mm and $\omega_c/2\pi=5.26\times 10^{14}$ Hz. The laser frequency $\omega_L/2\pi=2.82\times 10^{14}$ Hz.
\begin{figure}[!htb]
\begin{center}
\includegraphics[width=.4\textwidth]{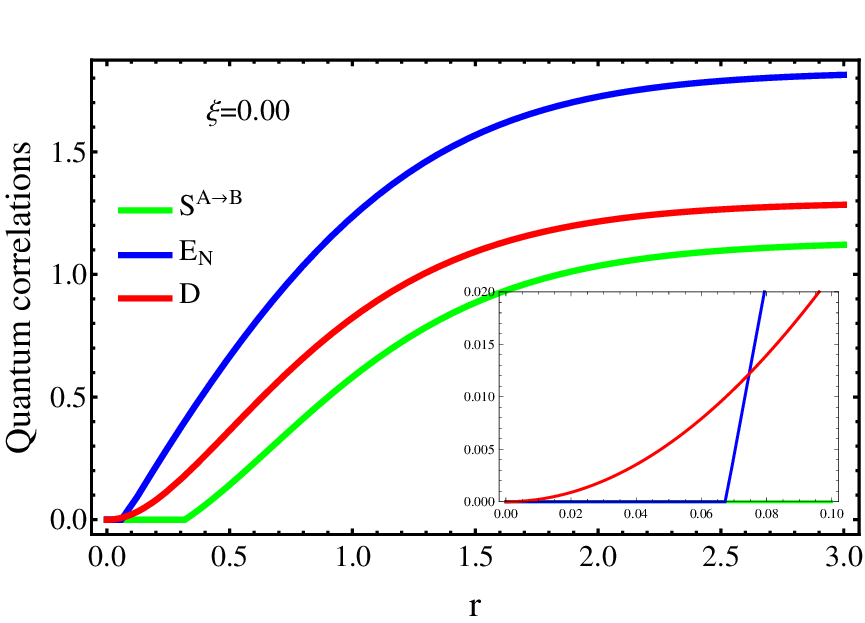}
\includegraphics[width=.4\textwidth]{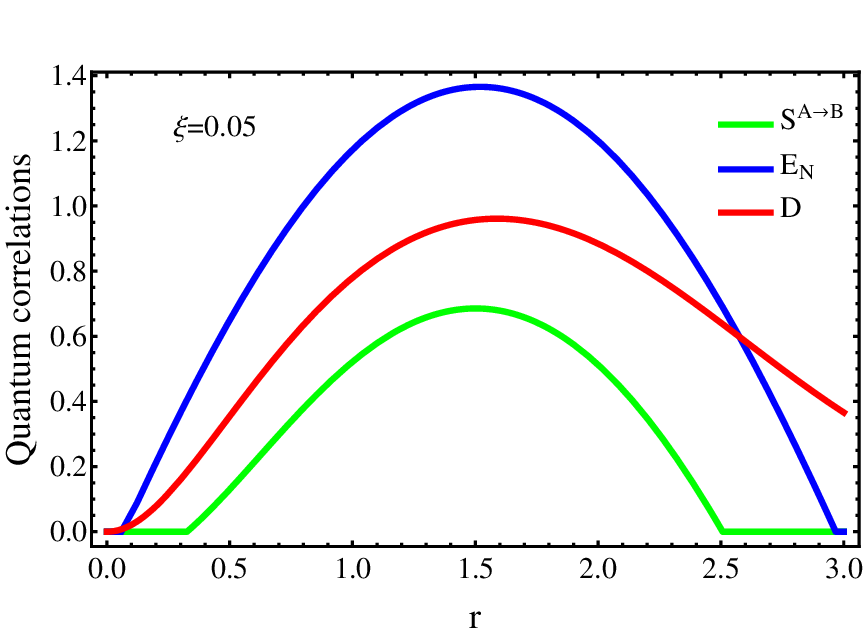}
\includegraphics[width=.4\textwidth]{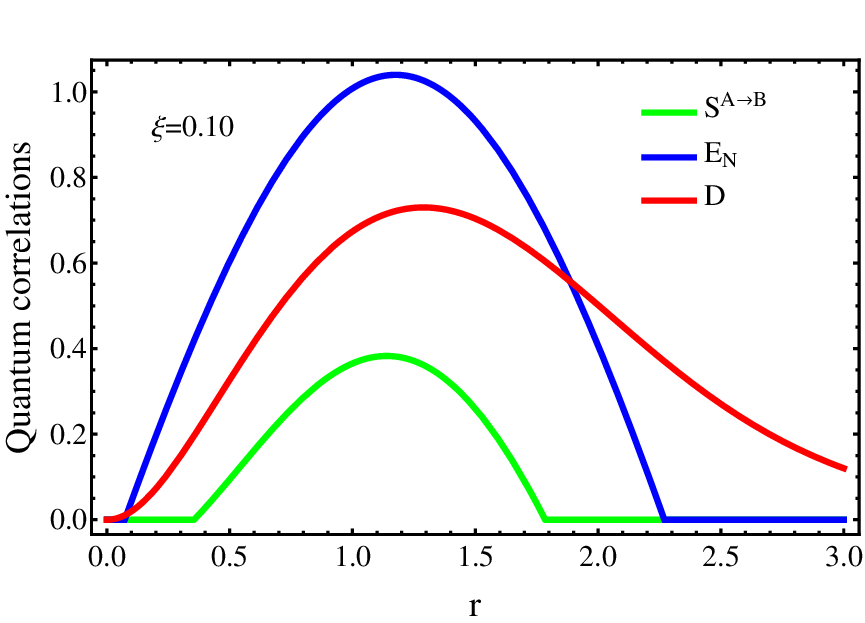}
\includegraphics[width=.4\textwidth]{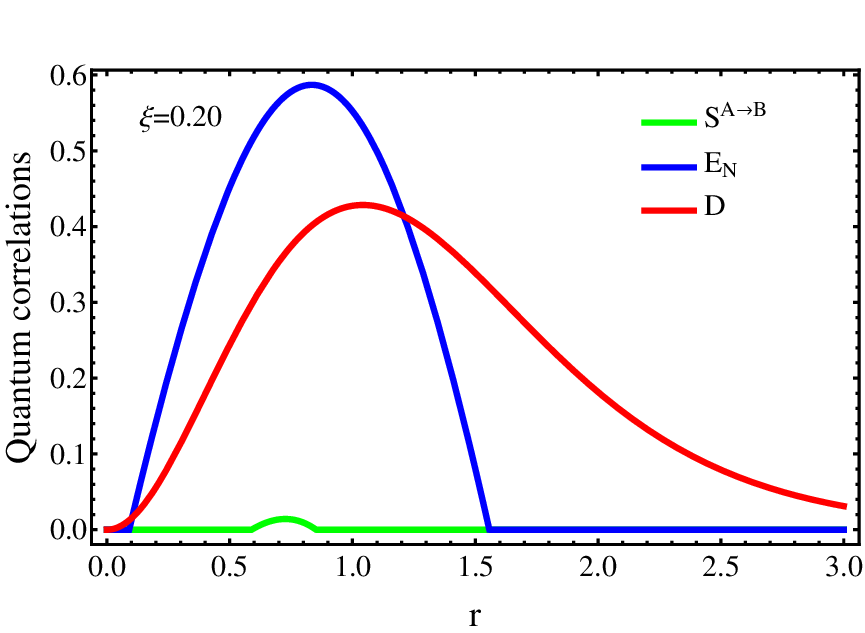}
\end{center}
\caption{Plots of the Quantum correlations (QCs), e.g. steering $S^{A\to B}$, logarithmic negativity $E_N$ and Gaussian quantum discord $\mathcal{D}$ of the two mechanical modes versus the squeezing parameter $r$ for various values of the the photon hopping strength $\xi$, with $\kappa=2\pi\times14000$ Hz and bath temperature of the mechanical modes $T = 0.1$ mK.}
\label{Nhoprxi}
\end{figure}

In Fig. \ref{Nhoprxi}, we plot the steering, entanglement and Gaussian quantum discord of mechanical modes against the squeezing parameter $r$ for different values of the photon hopping strength $\xi=\lambda/\kappa$. This figure shows that the generation of steering, entanglement and Gaussian quantum discord between movable mirrors need a minimum value of $r_{min}>0$. For a given $\xi$, the entanglement is achieved when $r>r_{min}$, as soon as $r_{min}$ increases with increasing values of $\xi$. So, the emergence of entanglement can be explained by the phenomenon of entanglement Sudden birth \cite{ZFicek2006}. We notice that when $\xi=0$ (i.e. in the absence of photon hopping interaction between the two cavities) steering, entanglement and Gaussian quantum discord increase monotonically with increase values of $r$ \cite{EASete2014,amaziougEPJD2018}. However, for a given value of $\xi>0$, steering, entanglement and Gaussian quantum discord increase when $r$ increases, and they decrease quickly once they achieve the maximum value (i.e. one can say that this happen by the resonance phenomenon between two movable mirrors). Indeed, the amount of steering, entanglement and Gaussian quantum discord decrease even when $r$ increases. This can be explained by the diminution of the effect of radiation pressure i.e. many photon optomechanical coupling $\mathcal{G}$ decreases (see also \cite{SBougouffa2019}). In addition, as seen in Fig. \ref{Nhoprxi}, the steering, entanglement and Gaussian quantum discord is affected by the photon hopping interaction. As expected, there exists a direct relationship between the generation of steering, entanglement and Gaussian quantum discord between mechanical modes with both photon hopping and squeezed light.\\
\begin{figure}[!htb]
\begin{center}
\includegraphics[width=.4\textwidth]{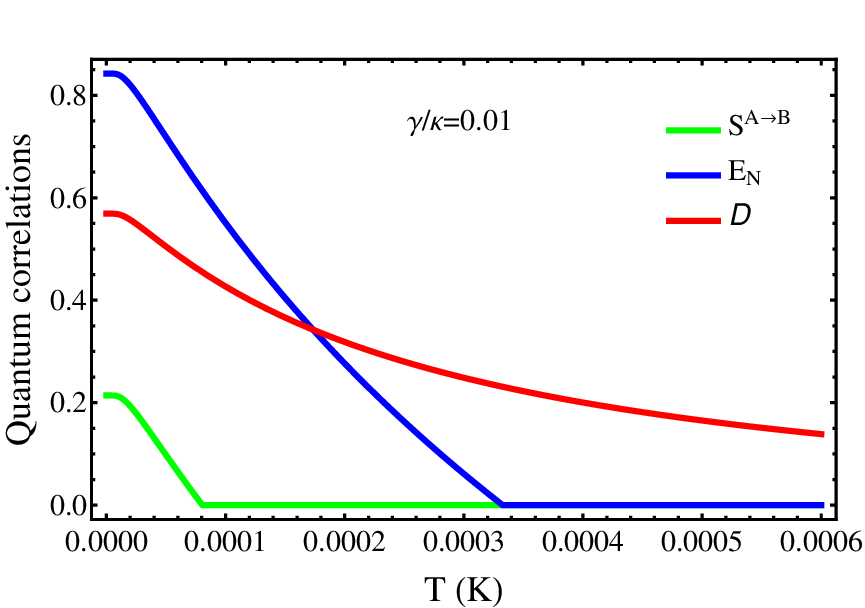}
\includegraphics[width=.4\textwidth]{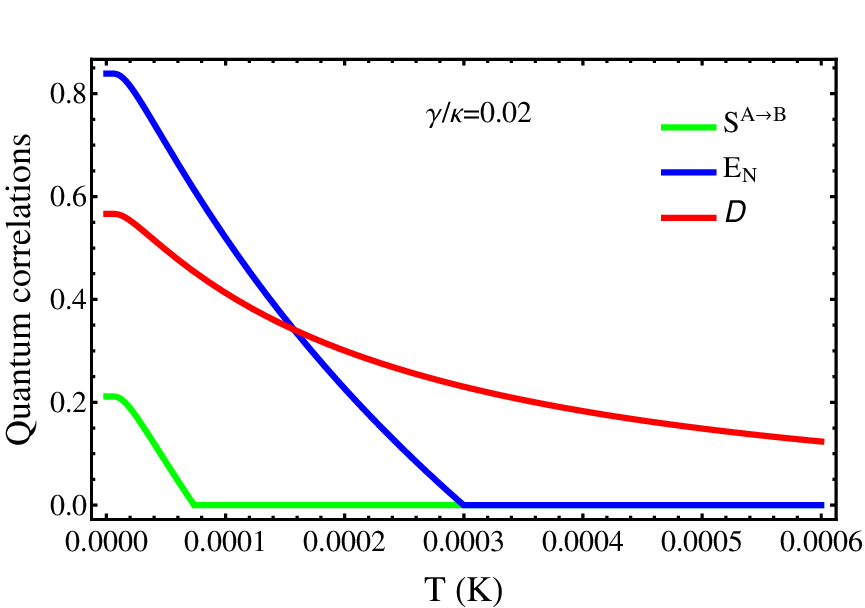}
\includegraphics[width=.4\textwidth]{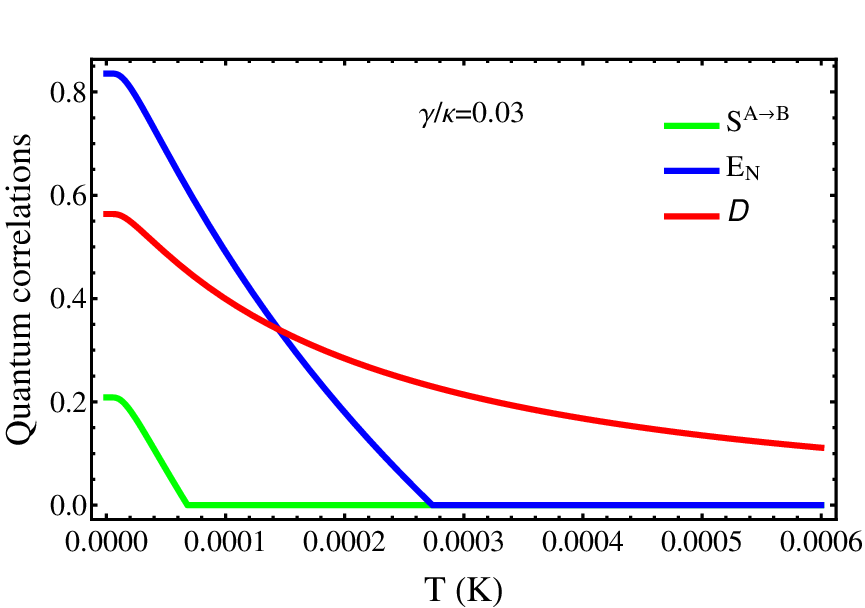}
\includegraphics[width=.4\textwidth]{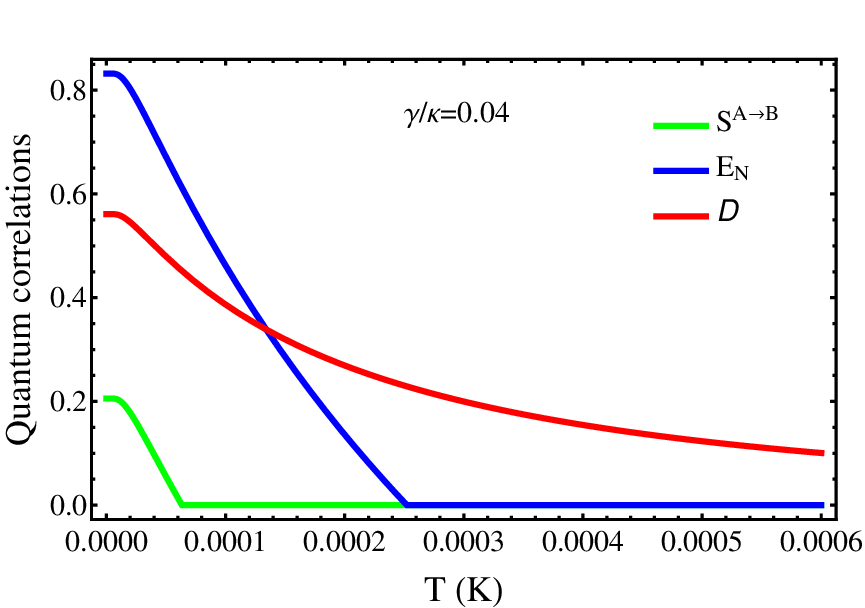}
\end{center}
\caption{Plots of the quantum correlations (QCs), e.g. steering $S^{A\to B}$, logarithmic negativity $E_N$ and Gaussian quantum discord $\mathcal{D}$ of the two movable mirrors versus bath temperature $T$ of the two movable mirrors for various values of $\gamma/\kappa$ for the photon hopping strength $\xi=0.2$, the squeezing parameter $r$=1 and the optomechanical cooperativity $\mathcal{C}=32.11$}
\label{NhopCxi}
\end{figure}

Fig. \ref{NhopCxi} shows the evolution the steering, entanglement and Gaussian quantum discord between two movable mirrors against bath temperature $T$ of the two movable mirrors for different values of $\gamma/\kappa$. 

We note on one hand that when $T$ increases,  the steering, entanglement and Gaussian quantum discord, between the mechanical modes,
decrease. This due by the phenomenon of entanglement decay under
the environment effects \cite{AAlQasimi2008}.. Furthermore, one notices that the vanishing values of the entanglement whereas steering and quantum are non zero.  This disappearance
of entanglement reflects its fragility versus the degradation effects induced by the environment. On the other hand the steering, entanglement and Gaussian quantum discord increases when $\gamma/\kappa$ decreases. This means that the transfer of quantum correlations is not efficient when dissipation rate of movable mirrors $\gamma$ is comparable to the cavity decay $\kappa$, i.e. when $\gamma$ increases thus the mechanical quality factor ($Q=\omega_M/\gamma$) is low and this leads to a less efficient entanglement transfer \cite{EASete2014}.

\begin{figure}[!htb]
\begin{center}
\includegraphics[width=.4\textwidth]{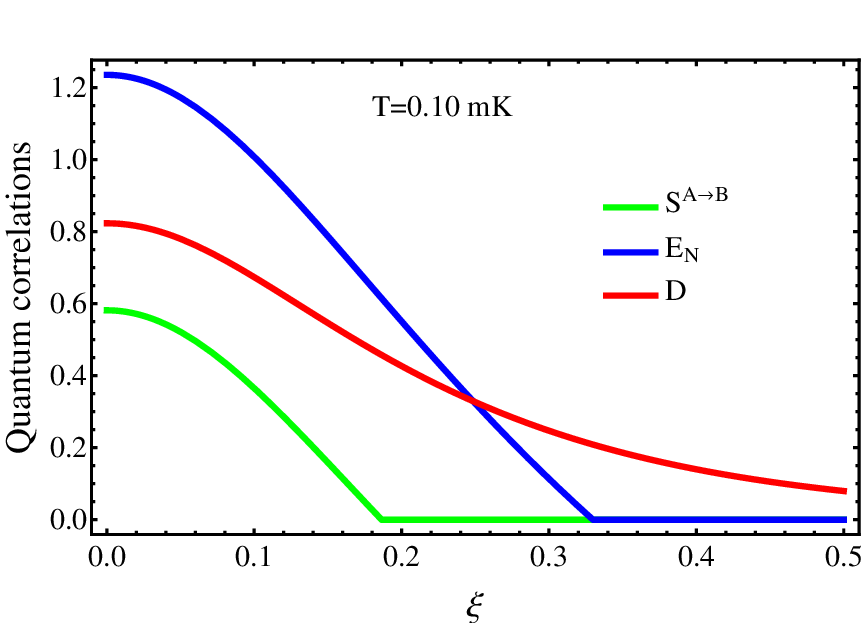}
\includegraphics[width=.4\textwidth]{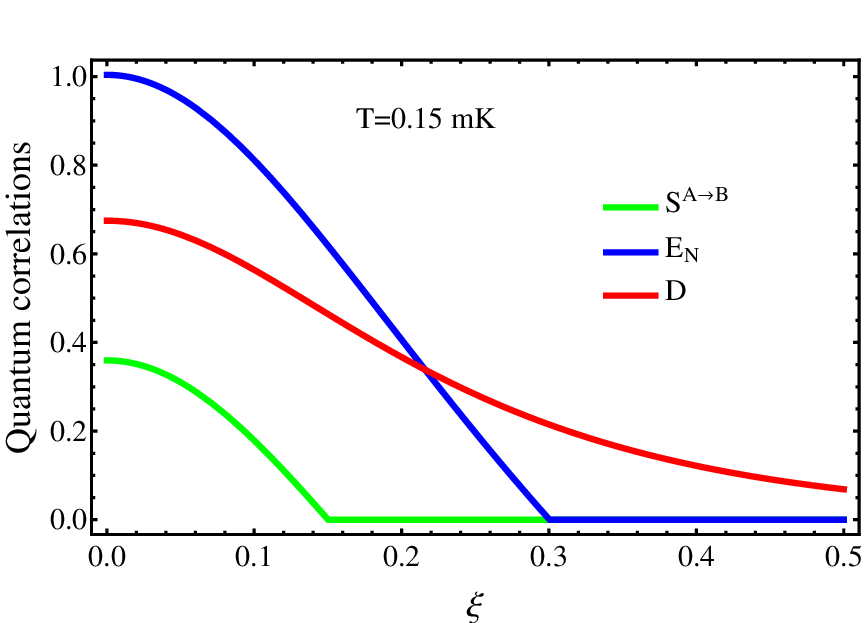}
\includegraphics[width=.4\textwidth]{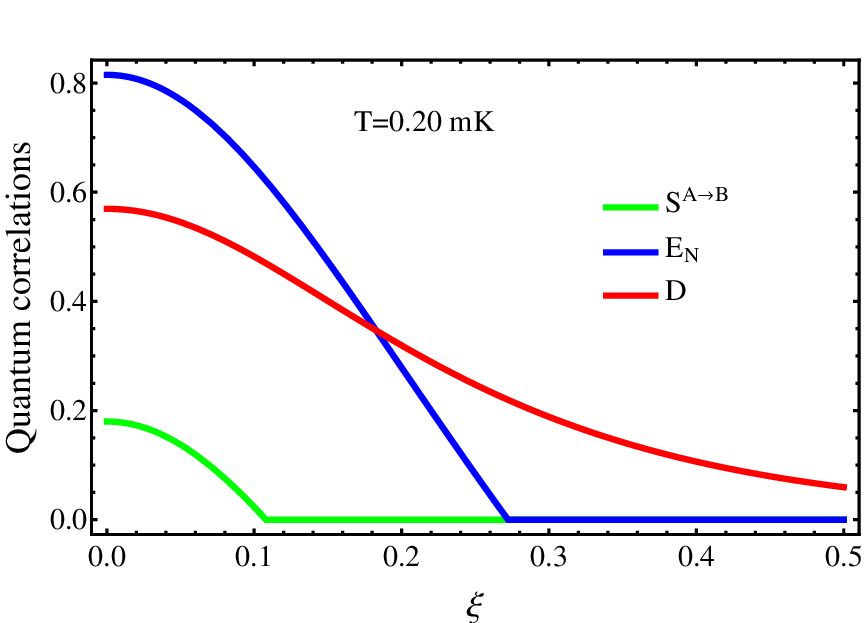}
\includegraphics[width=.4\textwidth]{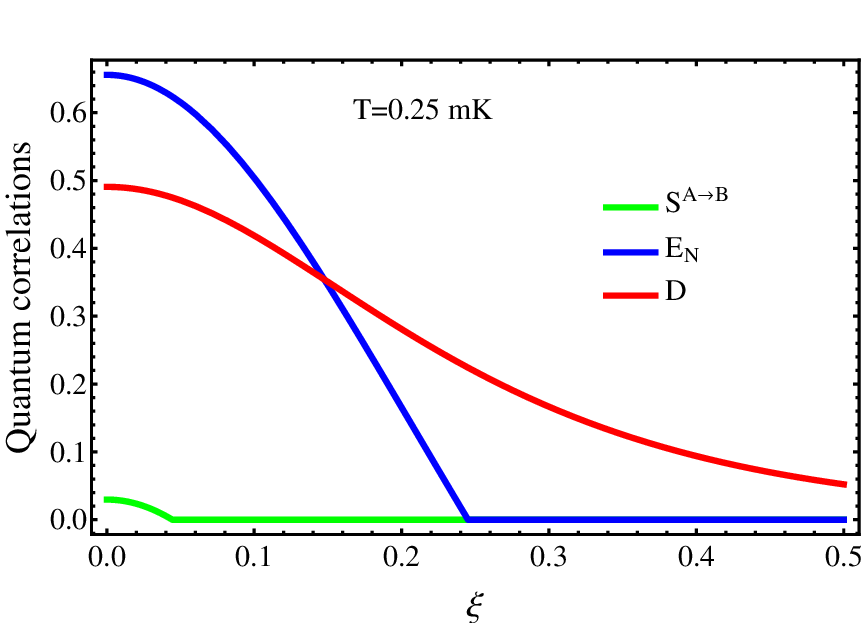}
\end{center}
\caption{Plots of the quantum correlations (QCs), e.g. steering $S^{A\to B}$, logarithmic negativity $E_N$ and Gaussian quantum discord $\mathcal{D}$ of the two movable mirrors versus the photon hopping strength $\xi$ for different values of bath temperature $T$, with $\kappa=2\pi\times14000$ Hz and squeezing parameter $r$=1}
\label{NhopTxi}
\end{figure}

In Fig. \ref{NhopTxi}, we plot the steering, entanglement and Gaussian quantum discord of two movable mirrors as a function of the photon hopping strength $\xi=\lambda/\kappa$ for various values of the temperature $T$ of the thermal bath of the movable mirrors. We notice, as this figure shows for a given $T$ steering, entanglement and Gaussian quantum discord decreases with $\xi$ increasing. This can be explained by the relation that exists between $\xi$ and $\mathcal{G}$. Furthermore, we notice that the quantumness of correlations decrease when $T$ increases this can be explained by the decoherence \cite{Zurek2003}.

The figures \ref{Nhoprxi}, \ref{NhopCxi} and \ref{NhopTxi} show that steering, entanglement and Gaussian quantum discord follow the same behavior. The gaussian quantum steering is bounded by entanglement (i.e. steerable state is a witness of entanglement, while the entangled mixed state is not always steerable). We have $S^{A\to B}=S^{B\to A}>0$ and logarithmic negativity $E_N>0$ is the witness of two-way steering, while for $S^{A\to B}=S^{B\to A}=0$ and $E_N>0$ the two movable mirrors are not steerable (i.e. no-way steering). Similarly, Gaussian quantum discord is more robust than entanglement as shown in figures \ref{Nhoprxi}, \ref{NhopCxi} and \ref{NhopTxi}, because when $0\leq \mathcal{D}<1$, i.e. the two movable mirrors are separable (if $E_N=0$) or entangled (if $E_N>0$). On the other hand when $\mathcal{D}>1$ (i.e. the two movable mirrors must be entangled) and we have $E_N>0$.

\section{Conclusion}

In this work, we have investigated the role of the  photon hopping process in enhancing the transfer of quantum correlations from
optical modes  and the mechanical modes. This work is largely inspired from the works related to quantum correlations in optomechanical systems and
it is motivated by the recently reported results in \cite{SBougouffa2017,SBougouffa2019} concerning the enhancement of non classical correlations transfer. We have employed and compared
three different non classical correlations (logarithmic negativity, quantum discord and quantum steering) and we have shown that quantum  correlations are enhanced the information is encoded 
 in squeezed states with non zero squeezing parameter  $r$ in the absence of  photon hopping (i.e.,  $\xi=0$) \cite{EASete2014}. This  is not always true when $\xi> 0$ as it has been reported in 
 \cite{SBougouffa2019}. In this picture, the mechanical modes are entangled when the squeezed parameter $r$ is no zero \cite{Jamal2015,amaziougEPJD2018}.
  By exploiting the recent experimental data \cite{SGroblacher2009}, we have found the existence of relationship of entanglement with the temperature $T$, the squeezing parameter $r$, photon hopping strength $\xi$ and the ratio $\gamma/\kappa$ involving the modes coupling parameters. These results are reported in Figs. (\ref{Nhoprxi}), (\ref{NhopCxi}) and (\ref{NhopTxi}). We have shown that steering, entanglement and discord decrease when the temperature increases (decoherence phenomenon) \cite{EASete2014,amaziougEPJD2018}. We have also noticed that the   robustness of mirror-mirror entanglement when the photon hopping strength is below its  asymptotic value $\xi_l$ i.e. entanglement between the mechanical modes persist when $\xi<\xi_l$ ($E_N=0$ for $\xi\geq\xi_l$) as showed in  Fig \ref{NhopTxi}.  We have found that Gaussian steering remains lower than entanglement, i.e. the steerable modes are strictly entangled but the entangled modes are not necessarily steerable. Finally, we have used the Gaussian quantum discord to quantify the quantum correlations, beyond entanglement, that can exist between two movable mirrors. In this way, it has been proved that Gaussian quantum discord  is a robust indicator of quantum correlations against temperatures environments. The hierarchy (entanglement, quantum steering, quantum discord ) of quantum correlations between two movable mirrors in optomechanical system can be realized in other quantum systems. We hope to report in this issue in a forthcoming work.
  
\textbf{Author contributions}

The authors contributed equally to this work.

\end{document}